\documentclass[%
 pra,
 amsmath,amssymb,
 reprint,%
]{revtex4-2}

\usepackage{amsmath,amsfonts}    
\usepackage{graphicx}   
\usepackage{verbatim}   
\usepackage{color,xcolor}      
\usepackage{subfigure}  
\usepackage{hyperref}   
\usepackage{textcomp}
\usepackage{bm}
\usepackage{bbm}
\usepackage{braket}

\graphicspath{{figs/}}

\draft 

\begin{document}


\title{Heralding on the Detection of Zero Photons} 



\author{C.M. Nunn}
\author{J.D. Franson}
\author{T.B. Pittman}
\affiliation{Department of Physics, University of Maryland Baltimore County, Baltimore, MD 21250, USA}


\date{\today}

\begin{abstract}
Although heralding signals in quantum optics experiments are typically based on the detection of exactly one photon, it has recently been theoretically shown that heralding based on the detection of zero photons can be useful in a number of quantum information applications. Here we experimentally investigate a technique for ``heralding on zero photons'' using conventional single-photon detectors. We illustrate how detector efficiency and dark count rates play a counterintuitive role in the ability to accurately detect zero photons, and use a variant of the Hong-Ou-Mandel interferometer to study the deleterious effects of limited detector efficiency when heralding on zero.
\end{abstract}


\maketitle 



\section{Introduction}\label{sec:intro}

Conditional measurements have a long and fruitful history in quantum optics \cite{kok_10} and quantum state engineering experiments \cite{dak_98-2}. The process often involves systems in which the detection of a single photon in one output mode is used to probabilistically herald the presence of a desired quantum state in a different output mode. A particularly powerful early example is the realization of a single-photon source based on Parametric Down Conversion (PDC) \cite{hm_86}. There, the detection of one member of a randomly emitted photon pair is used to herald the presence of the twin photon, which can then be used for subsequent applications \cite{pitt_02,mig_02}. More complex heralding examples include ideas ranging from photon subtraction \cite{our_06} and quantum scissors \cite{pegg_98}, to the realization of probabilistic entangling gates in the Linear Optics Quantum Computing (LOQC) paradigm \cite{klm_01}. Many of these applications have benefited from the recent advances in the development of photon-number resolving (PNR) detectors \cite{young_20} which enable heralding signals based on the detection of exactly 2, 3, or even larger numbers of photons \cite{thek_20,mattioli_15}.

\begin{figure}[t]
	\centering
	\includegraphics[scale=0.45,trim={210 190 210 190},clip]{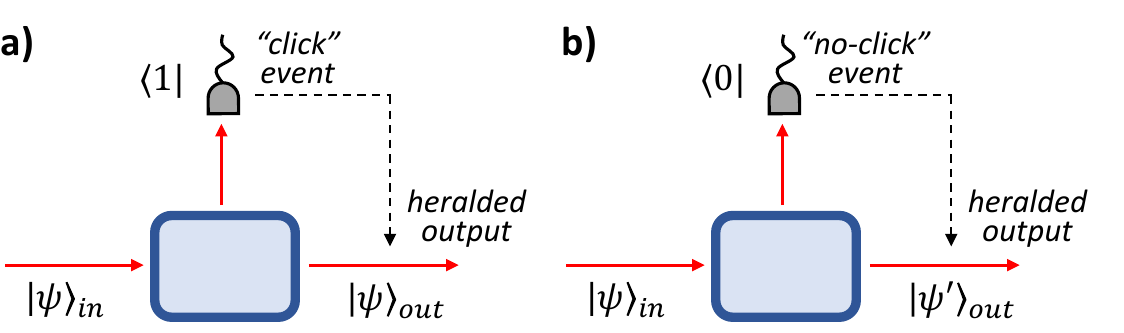}
	\caption{Comparison of quantum state engineering via (a) conventional heralding on the detection of exactly 1 photon, and (b) heralding on the detection of exactly zero photons. In both cases, a unique detection signal (a ``click" or ``no-click'' event) in one output mode is used to probabilistically herald the presence of a desired output state in the other output mode.}
	\label{fig:diag}
\end{figure}

Somewhat surprisingly, heralding signals based on the detection of zero photons are also useful. For example, ``heralding on zero'' is an essential feature of LOQC \cite{klm_01}, and forms the basis of noiseless attenuation for quantum communications \cite{mic_12,gag_14,ricky_17}. Figure 1 provides a simplified overview of the operational principle within these contexts. Here, an input state $|\psi\rangle_{in}$ enters a system that contains, for example, a carefully designed interferometric device with two output ports. In panel (a), the device is designed in such a way that the detection of exactly 1 photon in the upper output mode (a ``click'' event) heralds the presence of the desired state $|\psi\rangle_{out}$  in the other output mode. In panel (b), the device is designed so that the detection of exactly zero photons (a ``no-click" event) heralds a different desired output state  $|\psi^{\prime}\rangle_{out}$.

The core idea is that regardless of photon number, conditioning on the detection of a Fock state $|n\rangle$ in one mode can be a powerful tool for quantum state engineering \cite{dak_98-2}. Although extending this idea to $n=0$ is straightforward in theory, experimentally detecting zero photons presents a number of challenges~\cite{lita_08,lamp_14,helv_19} and motivates the need for effective ``heralding on zero" techniques~\cite{allevi_10,zhai_13,bog_17,hlou_17,magl_19,kat_20,sin_20}. In this paper, we explore the use of standard single-photon counting modules (SPCMs) for this unique application. In comparison with conventional heralding techniques based on the detection of 1 (or more) photons, the problematic roles of detector inefficiency and dark counts in realistic detectors are reversed. We begin by highlighting this idea with a simplified {\em gedankenexperiment}, and then experimentally demonstrate the effects through a unique signal that arises when ``heralding on zero'' in an otherwise conventional Hong-Ou-Mandel (HOM) interferometer~\cite{hom_87} fed with photon pairs from a pulsed PDC source.

\section{Overview}\label{sec:hoz}

\begin{figure*}[t]
	\includegraphics[scale=0.55,trim={25 140 25 130},clip]{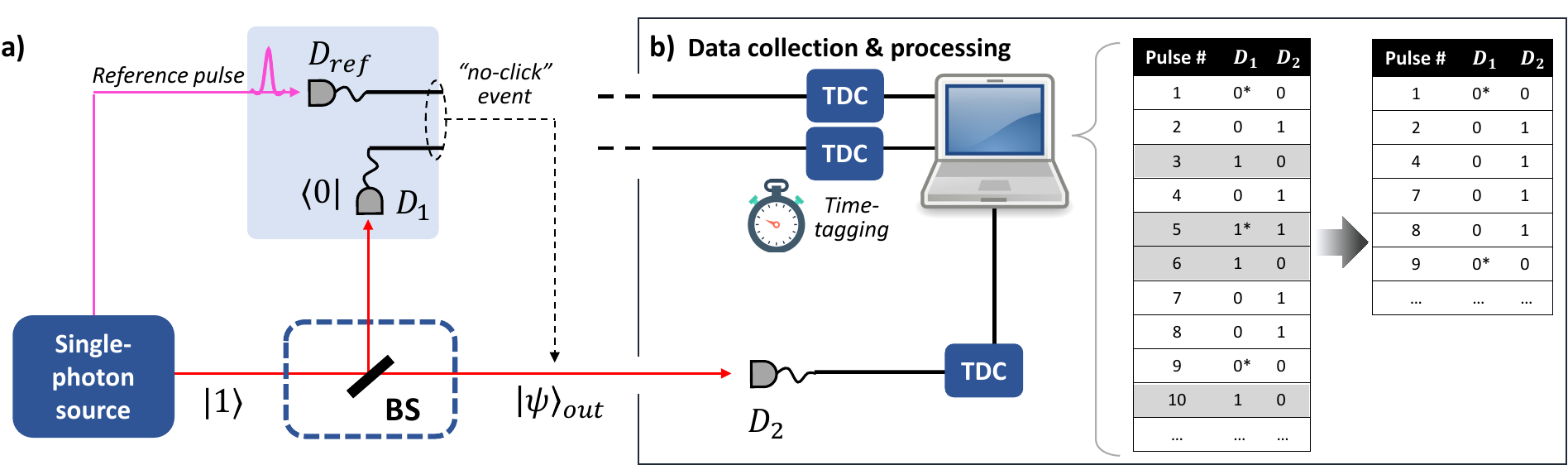}
	\caption{(a) Simple \textit{gedankenexperiment} using a pulsed single-photon source and 50/50 beamsplitter (BS) to highlight the role of detector inefficiency and dark counts when heralding on zero. Here, a detection at $D_{ref}$ combined with the absence of a detection signal from $D_1$ results in a ``no-click'' event, ideally heralding a single photon in the output. (b) Visualization of the measurement process used to verify the outcome in a realistic experimental setup using an auxiliary detector $D_2$, and time-to-digital converters (TDCs) for data collection and post-processing. $D_1$ dark counts and inefficiency errors are marked with asterisks, and negatively impact the heralded output state.}
	\label{fig:ex1}
\end{figure*}

To illustrate the basic technique of detecting zero photons with conventional single-photon detectors, we consider the simple \textit{gedankenexperiment} experiment shown in Figure \ref{fig:ex1}a. An input Fock state $\ket{1}$ is prepared with an idealized pulsed single-photon source (SPS), then sent into a 50/50 beamsplitter (BS). This single optical element serves as the general interferometric device first shown in Figure~\ref{fig:diag}b (the blue box). Reflected photons from the beamsplitter are monitored with a single-photon detector $D_1$. Additionally, each time the SPS emits a single photon, it also emits a strong optical reference pulse that is detected by an auxiliary reference detector $D_{ref}$ with perfect efficiency. As shown in the shaded box of Fig. 2(a), the detection of zero photons simply corresponds to a detection by $D_{ref}$, and the simultaneous absence of a detection by detector $D_{1}$. Crucially, the unaccompanied reference pulse enables a detectable signal (i.e., a ``no-click'' event). In this \textit{gedankenexperiment}, the ``no-click'' event heralds the presence of a single photon in the output port of the beamsplitter.

This simple example highlights the role of dark counts and detector inefficiency when heralding on zero. A dark count at $D_1$ alongside a signal from $D_{ref}$ will register as a ``click,'' leading us to discard the output even if the single photon was transmitted. This reduces the probability of success for our device. Detector inefficiency, meanwhile, will cause us to miss reflected photons and erroneously herald none in the output. This reduces the fidelity of states heralded by a ``no-click'' event. Significantly, the effects of dark counts and inefficiency are reversed from conventional heralding on single photons \cite{daur_12,silb_10}, so what makes $D_1$ a good ``zero photon detector'' may be defined by different criteria. For example, techniques which reduce losses at the cost of increased background noise may be beneficial in certain applications~\cite{gag_14}.

To quantify the effects of dark counts and detector inefficiency, we can measure the output state with an auxiliary detector $D_2$ and analyze the statistics of many trials, as shown in Figure \ref{fig:ex1}b. First, detection events from $D_{ref}$, $D_1$ and $D_2$ are processed by time-to-digital converters (TDCs) and stored as time tags relative to a master clock. This timing information allows us to reconstruct a sequence of ``click'' and ``no-click'' events alongside every output measurement \cite{sin_20}. The first table in Figure \ref{fig:ex1}b shows such a record with various $D_1$ detection errors marked with asterisks. The consequences of these errors appear in the \textit{subset} of data selected based on a ``no-click'' event at $D_1$, shown in the second table. The number of rows in this second table reflects the probability of success, and its output column should ideally contain only single photons. Dark counts at $D_1$ (e.g., Pulse \#5) are simply excluded from the subset of successful trials, reducing its size. Meanwhile, undetected photons at $D_1$ due to detector inefficiency (e.g., Pulses \#1 and \#9) are accepted as ``successful'' trials despite no photons being transmitted, corrupting the heralded output state. A more quantitative analysis of the effects of dark counts and detector inefficiency is shown in Appendix~\ref{app:A}.

In a realistic experiment with conventional silicon-based avalanche photodiodes (i.e., SPCMs), the $D_1$ dark count rates ($\sim10^2$ Hz) are much smaller than typical trial rates (often $10^4-10^7$ Hz). Thus, the change in probability of success will be negligible for many applications. In contrast, $D_1$ efficiency (typically $\sim50$\%) impacts a significant fraction of the trials and will drastically reduce the output fidelity in a realistic ``heralding on zero'' process. Consequently, the remainder of this paper will focus on the effects of $D_1$ inefficiency on the heralded output.

\section{Experimental Demonstration}\label{sec:hom}

\begin{figure}
	\includegraphics[scale=0.45,trim={70 180 500 120},clip]{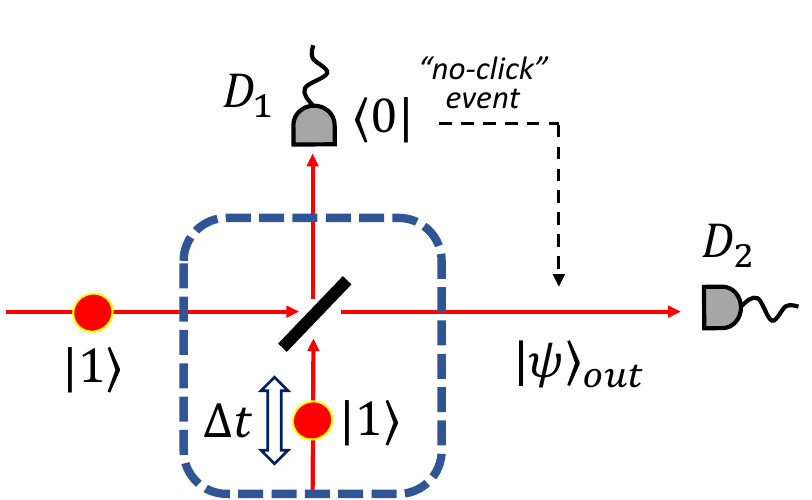}
	\caption{Conceptual illustration of heralding on zero in a modified Hong-Ou-Mandel (HOM) interferometer. Two identical photons (red circles) enter a 50/50 beamsplitter, and are made temporally distinguishable by a tunable delay $\Delta t$ in the lower input mode. As the delay is scanned through $\Delta t=0$, the \textit{probability} of heralding on zero doubles, resulting in a dramatic ``peak'' in the heralded $D_2$ event rate. However, the relative height of the peak rapidly degrades as a function of $D_1$ efficiency, providing a method for quantifying the role of detector efficiency in the heralding on zero process.}
	\label{fig:hom} 
\end{figure}

In order to experimentally demonstrate the role of limited detector efficiency in heralding on zero, we use a variant of the well-known Hong-Ou-Mandel (HOM) interferometer \cite{hom_87}. As pictured in Figure~\ref{fig:hom}, two photons are mixed at a 50/50 beamsplitter, resulting in four possible two-photon output amplitudes which we denote $RR$, $TT$, $RT$, and $TR$, with $R$ and $T$ implying single photon reflection and transmission, as usual~\cite{hom_87}. The photons are assumed to be indistinguishable, except for the temporal degree of freedom which is controlled by a relative delay $\Delta t$. When $\Delta t=0$, the beamsplitter statistics correspond to those of indistinguishable bosons, resulting in the suppression of the $RR$ and $TT$ amplitudes (i.e., bosonic ``bunching''~\cite{loudon_98}). Experimentally, this leads to the well-known HOM ``dip'' in coincident detections between $D_1$ and $D_2$ as $\Delta t$ is scanned through zero~\cite{hom_87}.

Instead of measuring coincidence counts while scanning $\Delta t$, here we herald on zero photons in $D_1$, and then study the single count rates in $D_2$ (see Figure~\ref{fig:hom}). In the idealized case, heralding on zero in $D_1$ simply means there are two photons in the output mode (i.e., headed to $D_2$)~\cite{giu_03,helv_19}. What is interesting, however, is that the \textit{probability} of this occurring depends on the value of $\Delta t$: when $\Delta t$ is large, heralding on zero in $D_1$ only occurs for one of the 4 possible two-photon amplitudes; when $\Delta t=0$, it occurs for one of only 2 possible two-photon amplitudes. Consequently, in an experiment with many repeated trials, the measurement process illustrated in Figure~\ref{fig:hom} should show a dramatic “peak” (i.e., doubling) in the heralded $D_2$ click rate as $\Delta t$ is scanned through zero time delay.

For our purposes, the key point is that the quality of this peak critically depends on the ability to effectively herald on zero, and the idealized doubling in counting rates rapidly degrades with $D_1$ efficiency, $\eta_1$.  Consequently, studying the relative peak height as a function of $\eta_1$ provides a robust metric for demonstrating the role of detector efficiency in heralding on zero.

\begin{figure*}[t]
	\includegraphics[scale=0.55,trim={30 145 30 160},clip]{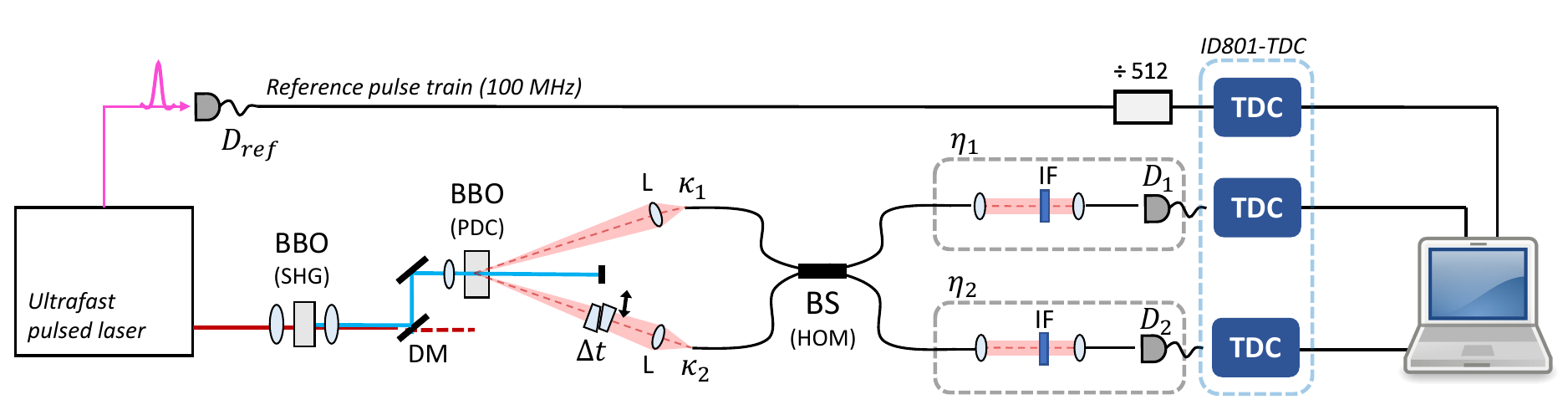}
	\caption{Experimental implementation of the modified HOM interferometer using photon pairs from a pulsed Parametric Down-Conversion (PDC) source, and a single-mode 3dB fiber coupler as the 50/50 beamsplitter (BS). The output of an ultrafast pulsed laser (100 MHz, 780 nm) undergoes Second Harmonic Generation (SHG) in order to produce a train of PDC pump pulses at 390 nm. The PDC crystal (BBO, Type I phase matching) produces photon pairs at 780 nm. Detectors $D_1$, $D_2$, and $D_{ref}$, as well as the TDCs are used to implement the measurement and data post-processing techniques illustrated in Figure~\ref{fig:ex1}b. DM-- dichroic mirror used to isolate UV pump pulses; L-- various lenses; $\Delta t$-- translating glass wedges; IF-- narrowband interference filters centered near 780 nm; ``$\div 512$''-- data frequency divider.}
	\label{fig:ex2}
\end{figure*}

This measurement can be accomplished with the same techniques described in Section~\ref{sec:hoz}, but with several additional considerations. First, we must account for the probability of ``no-click'' events given more than one photon at a non-PNR single-photon detector. This is given by $p(NC|n)=(1-\eta)^n$ for $n$ photons and negligible dark count probability \cite{resch_01,stev_13}. Here, the effective detector efficiency $\eta$ includes quantum efficiency, spatial mode overlap, and other forms of photon loss in the detection channel \cite{ware_04}. Next, our output measurement is no longer perfect, so the probability $p(NC|n)$ must be applied to both the heralding detector $D_1$ and the output detector $D_2$ (which is also inefficient and non-PNR). Lastly, our input state signaled by detector $D_{ref}$ (see Fig.~\ref{fig:ex1}) now corresponds to photon pairs derived from a probabilistic PDC source. We approximate the output of this source as a superposition of the vacuum and a single photon pair, ignoring higher-order terms~\cite{gk_04}:
\begin{equation}\label{eqn:in}
\ket{\psi}_{in}\approx\sqrt{1-\gamma}\ket{00}+\sqrt{\gamma}\ket{11}
\end{equation}
where $\gamma$ is the pair emission probability per pulse \mbox{($\sim10^{-4}$ in our experiment)}. The input state is further modified by finite coupling efficiencies $\kappa_1$ and $\kappa_2$, limited by the broadband pulsed PDC process \cite{pitt_05,bovino_03}, and separate from detector efficiencies $\eta_1$ and $\eta_2$ at the outputs.

Incorporating all of this into our analysis (see Appendix~\ref{app:B}), we obtain the probability of a $D_2$ click conditioned on a ``no-click'' at $D_1$:
\begin{eqnarray}
P(C_2|NC_1)\approx\frac{\gamma\eta_2^\prime}{4}\left[4-\eta_2^\prime-2\eta_1^\prime+\nu(2\eta_1^\prime-\eta_2^\prime)\right],\label{eqn:pcnc}
\end{eqnarray}
where $\eta_i^\prime\equiv \sqrt{\kappa_1\kappa_2}\eta_i$, and $\nu$ is a function of the time delay that captures the degree of indistinguishability of the two input photons. This parameter ranges from $\nu=1$ at $\Delta t=0$ to $\nu=0$ when $\Delta t$ is large. (For reference, the conventional HOM ``dip'' using the same apparatus would have the form $(1-\nu)$ as $\Delta t$ is scanned across zero.) The above expression also contains two approximations that are both valid in our experiments: first, that input coupling is roughly balanced $(\kappa_1\approx\kappa_2)$, and second, that $\gamma\ll 1$.

From Equation~\ref{eqn:pcnc} it can be seen that when $\eta_1^\prime=1$ (perfect efficiency), we find that $P(C_2|NC_1)\propto(1+\nu)$, indicating the high-visibility peak (doubling) in counts that does not depend on $\eta_2^\prime$. However, as $\eta_1^\prime$ decreases, the relative peak height also decreases monotonically until $\eta^\prime_1=0$. At this point, the heralding process is completely ineffective and we are simply observing the ordinary singles counting rate of $D_2$. This change of peak height as a function of $\eta_1^\prime$ is what we observe in our experiment.

The full experimental setup is shown in Figure~\ref{fig:ex2}. Our input state consists of down-converted photon pairs with center wavelength 780 nm, generated via degenerate Type-I PDC using a BBO crystal. The crystal is pumped with a 100MHz train of UV pulses ($\sim150$ fs duration, 390 nm center wavelength), derived from a frequency-doubled mode-locked fiber laser (Menlo Systems C-Fiber 780). The pairs are collected and focused into two single-mode fibers, with associated coupling efficiencies $\kappa_1$ and $\kappa_2$. The HOM interferometer consists of a 50-50 fiber coupler, as well as a pre-fiber time delay $\Delta t$ implemented with two translating glass wedges. After the HOM interferometer lie two detection channels; each consists of a free-space U-bench with 25-nm-bandwidth rectangular bandpass filters centered near 780 nm, before finally being coupled into multimode fibers and directed to SPCMs $D_1$ and $D_2$ (silicon avalanche photodiodes, Excelitas SPCM-AQ4C). Detection signals, including the 100 MHz mode-locking reference signal from $D_{ref}$, are recorded with TDCs with 81 ps timebin resolution (IDQuantique, model ID801). Since 100MHz exceeds the data bandwidth of our TDCs, the reference signal is buffered by frequency divider (Valon 3010a), so we tag one of every 512 pulses and reconstruct the pulse train in data processing.

The use of time tags and the data processing techniques of Fig.~\ref{fig:ex1}b offer several practical advantages. For example, by using an external reference signal, ``no-click'' events can be identified for the heralding detector, and the relevant output measurement statistics can be seen in post-selection~\cite{sin_20}. In addition, we can arbitrarily extend detector dead time in post-processing, mitigating the effects of afterpulsing \cite{ulu_00}. In our experiment, we ignore 5 pulses following a click in either detector to account for SPCM detector dead times of $\sim50$ ns. Furthermore, the reference pulse train (mode-locking signal) can be used as a \textit{virtual} gating signal for each free-running detector. We only accept counts within a 2 ns window after each reference pulse, discarding outlying dark counts. In our system, virtual gating reduces dark counts rates in each detector to a negligible rate of $\sim60$ per second, or $<10^{-6}$ per pulse. As a final note, these technical advantages are offset to some extent by significant storage requirements for reference pulse time tags, which can be mitigated with various techniques~\cite{wahl_20}.

\section{Results and Analysis}\label{sec:res}

\begin{figure}
    \centering
    \includegraphics[scale=0.58,trim={10 0 10 32},clip]{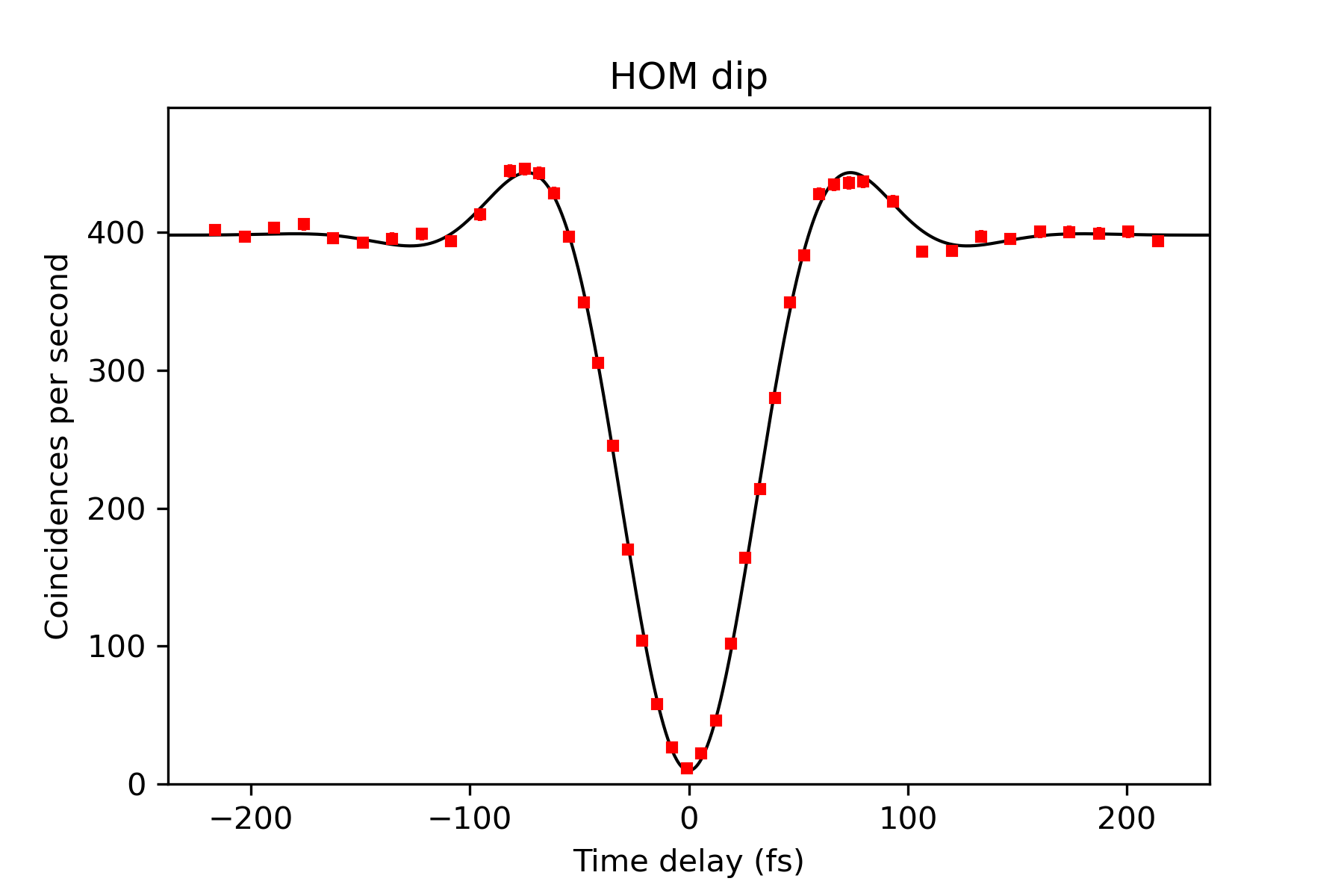}
    \caption{Alignment and testing of the setup using measurements of the conventional Hong-Ou-Mandel ``dip'' in coincidence counts between detectors $D_1$ and $D_2$. Experimental data points are in red. The solid black line is a least-squares fit using a simple model that takes into account the non-Gaussian transmission profiles of the narrowband interference filters. The model gives a HOM dip visibility of $(97.5\pm0.6)\%$, indicating a high degree of indistinguishability.}
    \label{fig:dip}
\end{figure}

We first align and optimize our apparatus by performing conventional Hong-Ou-Mandel measurements of the coincidence counting rate between $D_1$ and $D_2$ as a function of $\Delta t$. The optimized results are shown in Figure~\ref{fig:dip}, where each data point is displayed as an average counting rate, calculated from 20 seconds of stored time tags and a coincidence window of 2 ns. The results show a high-quality HOM ``dip'' with a visibility of $(97.5\pm0.6)\%$. This provides a measure of the indistinguishability of the photons and, importantly, an experimental upper bound for $\nu$ of $\nu_{\max}=0.975$.

\begin{figure}
    \centering
    \includegraphics[scale=0.55,trim={280 0 250 0},clip]{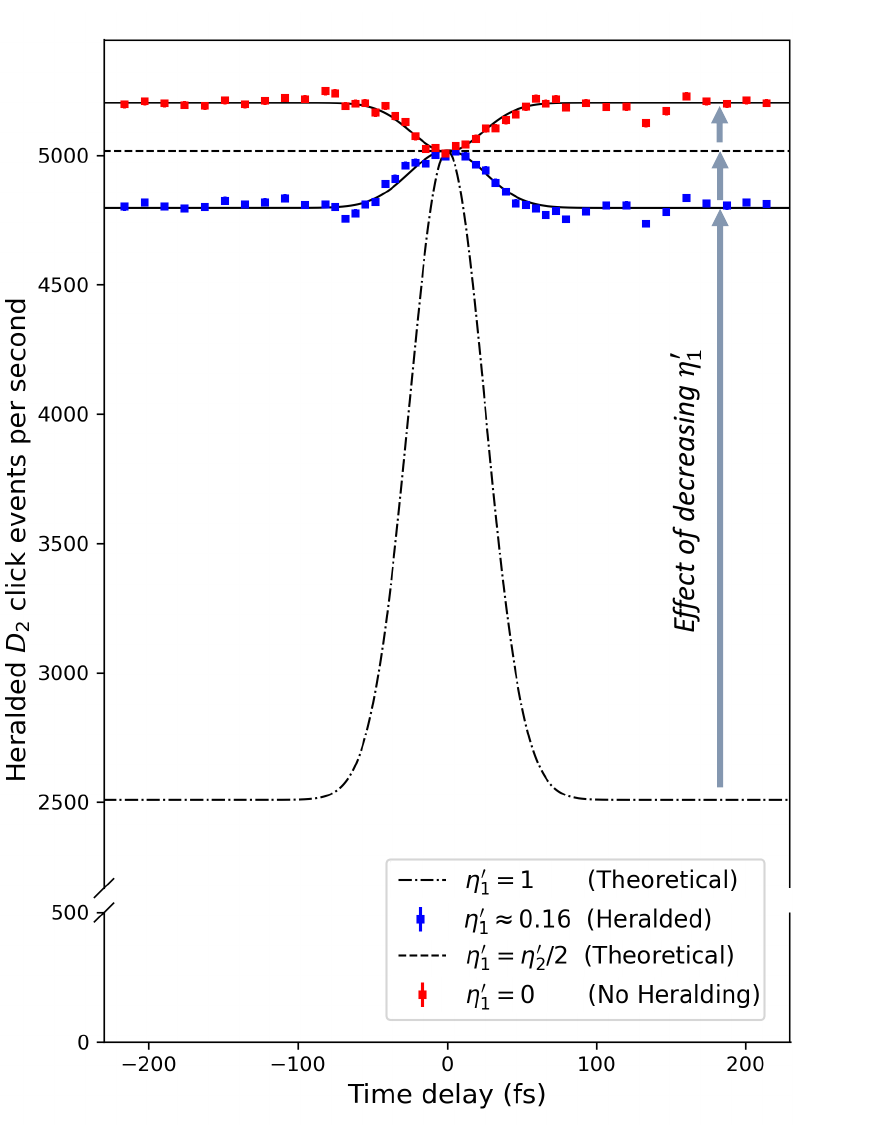}
    \caption{Demonstration of the deleterious effects of detector inefficiency when heralding on zero in the HOM interferometer. The 4 curves show the measured (or predicted) click rates at detector $D_2$ for different values of heralding detector efficiency $\eta_1^\prime$. For perfect efficiency $\eta_1^\prime=1$ (lowest curve), the heralded click rate doubles when $\Delta t=0$, resulting in a dramatic ``peak'' in the predicted curve. As $\eta_1^\prime$ is decreased, this effective peak height is reduced. The blue (lower) data points are experimental measurements taken with $\eta_1^\prime\approx0.16$. As $\eta_1^\prime$ is further reduced to the value $\eta_1^\prime=\eta_2^\prime/2$, the peak flattens (dashed line), and then transitions to a ``dip'' when $\eta_1^\prime<\eta_2^\prime/2$. The red (upper) data points correspond to the limiting case $\eta_1^\prime=0$ (i.e., no heralding). In all cases, output detector efficiency is fixed at the experimentally determined value $\eta_2^\prime\approx0.15$.}
    \label{fig:res}
\end{figure}

The main results of our ``heralding on zero'' study are summarized in Figure~\ref{fig:res}. The two experimental curves (blue and red data points) are derived from the same recorded time tags used to produce the results of Figure~\ref{fig:dip}. For convenience, we convert the probability per pulse, $P(C_2|NC_1)$ of Eqn~\ref{eqn:pcnc}, into heralded $D_2$ clicks per second using the repetition rate of the experiment. The data shows how the size of the expected ``peak'' in the heralded $D_2$ click rate is reduced with decreasing $D_1$ efficiency $\eta_1^\prime$.

For reference, the idealized case of $\eta_1^\prime=1$ is shown by the dashed theory curve in the lower part of Figure~\ref{fig:res}. This is simply the prediction given by Equation~\ref{eqn:pcnc} with $\gamma$ and $\eta_2^\prime$ experimentally determined. When $\Delta t=0$, the peak of the dashed theory curve shows the ideal doubling in click rate.

The blue data points are our experimentally measured values of $P(C_2|NC_1)$ with a low value of $\eta_1^\prime$. The idealized doubling in peak height is significantly reduced by raising the count rates at large time delays (the ``wings'') relative to the count rates at the central point when $\Delta t=0$. We quantify this reduction by taking the ratio of Eqn.~\ref{eqn:pcnc} at zero and large time delays (Appendix~\ref{app:B}), which we call the center-to-wings ratio $\xi$:
\begin{equation}\label{eqn:cwr}
    \xi\approx1+\nu_{\max}\left(\frac{2\eta_1^\prime-\eta_2^\prime}{4-2\eta_1^\prime-\eta_2^\prime}\right)
\end{equation}
In the ideal case $\eta_1^\prime=1$, the above simplifies to $1+\nu_{\max}$, but decreases monotonically with heralding detector efficiency. As $\eta_1^\prime$ worsens and count rates on the wings continue to rise, eventually one actually expects a relative ``dip'' in counts for zero time delay, $\xi<1$. This occurs in the worst case $\eta_1^\prime=0$, where heralding on zero is completely ineffective and admits all possible states into the output. This limiting case of no heralding is shown by the red data points in Figure~\ref{fig:res}, and corresponds to the dip in ordinary singles counting rates first observed by Resch et al.~\cite{resch_01} and Kim et al.~\cite{kim_03}.

For simplicity, we use basic Gaussian fits to our experimental data and find $\xi$ values of approximately $1.05$ and $0.96$ for our peak (blue) and dip (red), which correspond to effective detector efficiencies $\eta_1^\prime\approx0.16$ and $\eta_2^\prime\approx0.15$. This is consistent with the nominal detection efficiencies of our silicon avalanche photodiodes ($\sim50\%$ at 780 nm), combined with limited coupling efficiencies ($\kappa_1$ and $\kappa_2$ $\sim50$\%)~\cite{pitt_05,bovino_03}, and U-bench transmission ($\sim 65\%$). The theoretical curves in Figure~\ref{fig:res} are calculated using the same value of $\eta_2^\prime$ while varying heralding detector efficiency~\footnote{In practice, our ability to fix one effective detector efficiency while varying the other is limited, since they both depend on $\kappa_1$ and $\kappa_2$. However, it is more instructive to consider the efficiencies as independent, as if all losses occur in the detection channels and not before the HOM interferometer. Our analysis applies in either case}.

In Figure~\ref{fig:res}, both the ``peak'' (blue) and ``dip'' (red) results appear to share the same value at zero time delay. This is due to the fact that our model for heralded statistics in Eqn.~\ref{eqn:pcnc} is equivalent to the coincidence counting rate subtracted from the ordinary singles counting rates (see Appendix~\ref{app:B}) \footnote{Another explanation for this fixed point is more fundamental: including all higher-order terms from the PDC process, at $\Delta t=0$ the outputs of the beamsplitter should ideally be disentangled and thus heralding on $D_1$ should not affect the state in the other output mode \cite{kim_02}. It is only by comparison to nonzero time delays that we gain information about the heralding capability of detector $D_1$.}. Thus the difference between red and blue data points is nearly zero at zero time delay, and matches the experimental coincidence counting rate of Figure~\ref{fig:dip}. Additionally, since $\xi$ evolves continuously with heralding efficiency, there is a value of $\eta_1^\prime$ at which the interference pattern transitions from a ``peak'' to a ``dip.'' This condition is shown by the flat dashed line in Figure~\ref{fig:res}. By inspection of Eqn.~\ref{eqn:cwr}, this occurs when $\eta_1^\prime=\eta_2^\prime/2$. At this point, a heralded $D_2$ ``click'' event is equally likely to result from ``bunched'' ($TR$ or $RT$) or ``antibunched'' ($TT$ or $RR$) two-photon amplitudes, and so interference no longer appears in this measurement.

\begin{figure}
    \centering
     \includegraphics[scale=0.40,trim={185 70 180 80},clip]{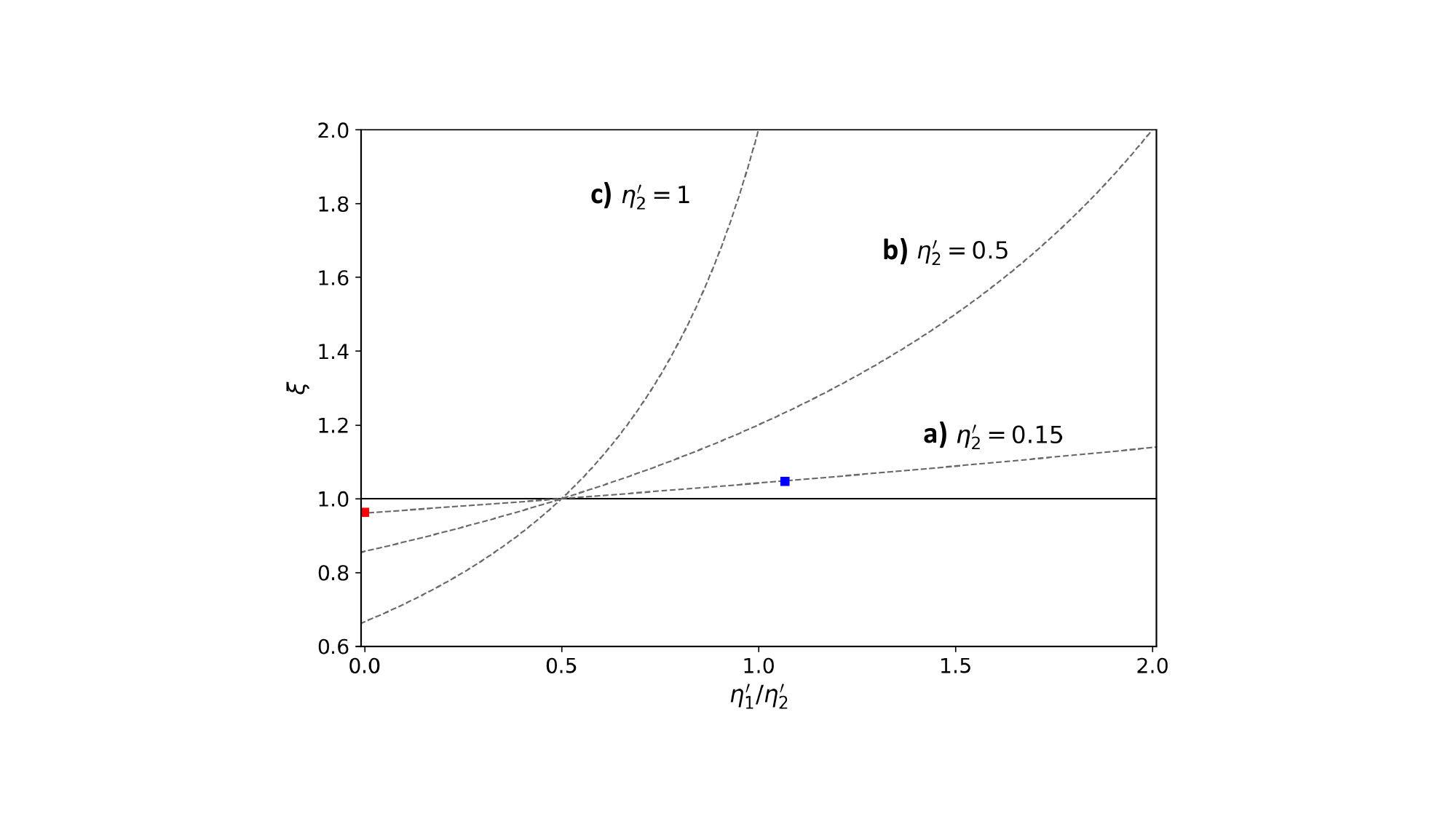}
        \caption{Theoretical plots of center-to-wings ratio $\xi$ as a function of $\eta_1^\prime/\eta_2^\prime$, shown for three values of output detector efficiency: a) the experimental value $\eta_2^\prime=0.15$, b) an intermediate efficiency $\eta_2^\prime=0.5$ and c) the ideal case $\eta_2^\prime=1$. Interestingly, no interference can be observed (i.e., $\xi=1$) when $\eta_1^\prime/\eta_2^\prime=1/2$, regardless of the actual detector efficiency values. When $\eta_1^\prime/\eta_2^\prime>1/2$, dramatic ``peaks'' ($\xi>1$) can be observed in the $D_2$ click rate with a maximum of $\xi=2$ for the ideal case of $\eta_1^\prime=1$ and $\nu_{max}=1$. When $\eta_1^\prime/\eta_2^\prime<1/2$, the ``peak'' transitions to a ``dip'' ($\xi<1$), with a minimum value of $\xi=2/3$ when $\eta_2^\prime=1$ and $\eta_1^\prime=0$ (i.e., equivalent to no heralding). For reference, the two points (red and blue) on curve (a) correspond to observed data in Figure~\ref{fig:res}.}
    \label{fig:curve}
\end{figure}

This transition is further theoretically explored in Figure~\ref{fig:curve}, where $\xi$ is plotted as a function of the ratio $\eta_1^\prime/\eta_2^\prime$ for a wide range of efficiencies extending beyond our experimental capabilities. For simplicity, here we also assume $\nu_{max}=1$. Three curves are shown for different values of output efficiency $\eta_2^\prime$, and all of them pass through the non-interference condition point $\xi=1$ when $\eta_1^\prime=\eta_2^\prime/2$. Above this threshold, we observe a peak in heralded $D_2$ click events. The upper bound of $\xi=2$ is achieved for perfect heralding efficiency $\eta_1^\prime=1$. Below the threshold, the efficiency of the output detector becomes most important as heralding efficiency drops to zero. Although the average photon flux at detector $D_2$ remains constant with no heralding ($\eta_1^\prime=0$), the lack of PNR results in a dip in singles rates. For the extreme case $(\eta_1^\prime,\eta_2^\prime)=(0,1)$, we see the largest possible dip of $\xi= 2/3$. This lower bound can be explained with the two-photon amplitude model: for large $\Delta t$, 3 of 4 possibilities send at least one photon to detector $D_2$; for $\Delta t = 0$, this reduces to 1 of 2 possibilities, yielding the above ratio. This dramatic $D_2$ singles rate ``dip'' is present in all conventional HOM experiments with SPCMs, but is significantly reduced by limited $D_2$ efficiency and other losses~\cite{resch_01,kim_03,lee_06}.

\section{Conclusions}\label{sec:con}

While the concept of heralding on the detection of zero photons is fairly straightforward, the implementation with current technology presents a number of challenges. We have investigated a basic approach that combines the detection of a reference (timing) signal with the simultaneous \textit{absence of a detection} in a commercial SPCM (silicon-based avalanche photodiodes). This combination successfully registers a ``no-click'' event that can then be used to actively herald the desired output state in various applications.

In this approach, the role of dark counts in the heralding detector merely reduces the probability of success~\cite{gag_14}. In contrast, the role of detector inefficiency severely reduces the fidelity of the heralded output states. In some sense, this makes experimental heralding on zero much more difficult than conventional heralding on one photon, where these roles are reversed.

We experimentally quantified the effects of detector inefficiency in heralding on zero using a variation of the HOM interferometer as a test system. The results showed a rapid deterioration of the relevant quantum interference effects as the heralding detector efficiency is reduced. Extensions of this demonstration to practical systems in quantum communications~\cite{mic_12,gag_14} and quantum state generation~\cite{ye_19,mik_19} support the need for current research efforts in increasing detector efficiency~\cite{eis_11,zadeh_21}. This is particularly relevant for more complex scenarios, such as LOQC applications that may require simultaneous heralding on zero in multiple heralding channels.

\begin{acknowledgments}
This work was supported by the National Science Foundation under grant number PHY-2013464.
\end{acknowledgments}

\appendix

\section{Heralding on Zero Statistics}\label{app:A}

Here we present a quantitative analysis to justify the conclusions of Section~\ref{sec:hoz} regarding detector efficiency and dark counts. We go back to the more general case of Figure~\ref{fig:diag}b, with arbitrary input states. We start with the probability that a non-PNR detector $D_1$ will \textit{not} click given exactly $n$ photons enter the detection channel \cite{resch_01,stev_13}:
\begin{equation}\label{eqn:p0n}
    p(NC|n)=(1-d)(1-\eta)^n,
\end{equation}
where $d$ is the probability of a dark count, assumed to be independent of $n$, and $\eta$ is the heralding detector efficiency. The probability of success $P_s$ is simply the total likelihood of a no-click event, $P(NC)$. Assuming a probability $P_n$ of $n$ photons appearing in the heralding mode:
\begin{equation}\label{eqn:ps}
    P_s=\sum_{n=0}^\infty(1-d)(1-\eta)^n P_n
\end{equation}
Thus, the probability of success decreases linearly with dark counts. However, only detector efficiency affects the quality (i.e., fidelity) of the heralded output. From Bayes' rule, we can calculate the probability of heralding the desired state $\ket{\psi}_{out}$ given a no-click event:
\begin{align}\label{eqn:pout}
    P(\psi_{out}|NC)&=\frac{P(NC|\psi_{out})\cdot P(\psi_{out})}{P(NC)}
    =\frac{(1-d)\cdot P_0}{P_s} \nonumber \\
    &=\frac{P_0}{\sum_{n=0}^\infty(1-\eta)^n P_n}
\end{align}
With unit detector efficiency $\eta=1$, we see that Eqn.~\ref{eqn:pout} is also unity, meaning a no-click event always heralds the desired output state. As $\eta$ decreases, however, this probability decreases as no-click events herald mixed states with added noise.

For example, in the single-photon \textit{gedankenexperiment} of Figure~\ref{fig:ex1}a, using a 50/50 beamsplitter, we have $P_0=P_1=1/2$. Thus, Eqn.~\ref{eqn:ps} gives \mbox{$P_s=(1-d)(2-\eta_1)/2$} and Eqn.~\ref{eqn:pout} gives $P(\psi_{out}|NC_1)=1/(2-\eta_1)$.

In this idealized setup, the degraded output state can be adequately characterized with the counting rates of one single-photon detector, $D_2$, accounting for additional losses. For a realistic single-photon source, which may also include small, undesired zero-photon and two-photon contributions, this would be better accomplished by a $g^{(2)}$ measurement in the output mode~\cite{silb_10}.

\section{HOM Heralded Click-rate Calculations}\label{app:B}
Including the coupling efficiencies $\kappa_1$ and $\kappa_2$, the probabilities $P(m,n)$ of finding $m$ and $n$ photons in the two HOM interferometer outputs are given by:
\begin{align*}
    P(1,1)&=\frac{\gamma\kappa_1\kappa_2}{2}(1-\nu) \\
    P(1,0)=P(0,1)&=\frac{\gamma}{2}\left[\kappa_1(1-\kappa_2)+(1-\kappa_1)\kappa_2\right] \\
    P(2,0)=P(0,2)&=\frac{\gamma\kappa_1\kappa_2}{4}(1+\nu)
\end{align*}
By applying Eqn.~\ref{eqn:p0n} and neglecting dark counts, we can obtain the probability of a click at either detector (singles), and a joint click event (coincidence):
\begin{eqnarray}
    P(C_i)= \frac{\gamma\eta_i\Tilde{\kappa}}{4}\left[\frac{4\Bar{\kappa}}{\Tilde{\kappa}}-\left(1+\nu\right)\eta_i\Tilde{\kappa}\right] \label{eqn:pp1} \\
    P(C_1\land C_2)= \frac{\gamma\eta_1\eta_2\Tilde{\kappa}^2}{2}\left(1-\nu\right)\label{eqn:pcoin}
\end{eqnarray}
where the input coupling efficiencies have been combined into $\Bar{\kappa}$ and $\Tilde{\kappa}$, the arithmetic and geometric mean, respectively. As long as the coupling efficiencies are roughly equal, we can approximate that $\Bar{\kappa}=\Tilde{\kappa}$.

Next, we find the probability of a $D_2$ click conditioned on a $D_1$ no-click in terms of Eqns.~\ref{eqn:pp1} and \ref{eqn:pcoin}, once again using Bayes' rule:
\begin{equation}\label{eqn:pclick}
    P(C_2|NC_1)=\frac{P(C_2)-P(C_1 \land C_2)}{1-P(C_1)}
\end{equation}
With small pair probability $\gamma\ll 1$, the denominator (i.e., the probability of success) can be taken as unity. This last approximation gives us Eqn.~\ref{eqn:pcnc}, which is equal to $P(C_2)-P(C_1\land C_2)$; this is simply the coincidence counting rate subtracted from the $D_2$ singles counting rate.

The center-to-wings ratio $\xi$ of Eqn.~\ref{eqn:cwr} is defined:
\begin{equation}\label{eqn:cwr2}
	\xi\equiv\frac{P(C_2|NC_1)|_{\nu=\nu_{max}}}{P(C_2|NC_1)|_{\nu=0}}\approx1+\nu_{\max}\left(\frac{2\frac{\eta_1^\prime}{\eta_2^\prime}-1}{\frac{4}{\eta_2^\prime}-2\frac{\eta_1^\prime}{\eta_2^\prime}-1}\right),
\end{equation}
where the above form illustrates the dependence on efficiency mismatch $\eta_1^\prime/\eta_2^\prime$, illustrated in Fig.~\ref{fig:curve}.

Note that the approximation in Eqn.~\ref{eqn:cwr2} only holds for $\gamma\ll 1$ and $\kappa_1\approx\kappa_2$. When these conditions are not met, the presence of higher-order terms (multiple pairs) in the PDC process and largely asymmetric coupling losses can affect both $\xi$ and the conventional HOM dip visibility in these types of experiments~\cite{gutr_15}.

\end{document}